\documentclass[sigconf]{acmart}
\AtBeginDocument{%
  }


\makeatletter
\expandafter\let\csname not=\endcsname\relax
\expandafter\let\csname not<\endcsname\relax
\expandafter\let\csname not>\endcsname\relax
\makeatother
\usepackage{libertinus}
\usepackage{bbding}
\usepackage{adjustbox}
\usepackage{multirow}
\usepackage{float}
\usepackage{graphicx}

\copyrightyear{2025}
\acmYear{2025}
\setcopyright{acmlicensed}\acmConference[MM '25]{Proceedings of the 33rd ACM International Conference on Multimedia}{October 27--31, 2025}{Dublin, Ireland}
\acmBooktitle{Proceedings of the 33rd ACM International Conference on Multimedia (MM '25), October 27--31, 2025, Dublin, Ireland}
\acmDOI{10.1145/3746027.3758260}
\acmISBN{979-8-4007-2035-2/2025/10}

\begin{document}
\begin{sloppypar}

\title{\textbf{AudioSet-R}: A Refined AudioSet with Multi-Stage LLM Label Reannotation}

\author{Yulin Sun}
\orcid{0009-0001-3717-2223}
\authornote{Both authors contributed equally to the paper}
\author{Qisheng Xu}
\authornotemark[1]
\affiliation{%
  \institution{College of Computer Science and Technology, National University of Defense Technology}
  \city{Changsha}
  \state{Hunan}
  \country{China}
}
\email{sunyulin@nudt.edu.cn}
\email{qishengxu@nudt.edu.cn}

\author{Yi Su}
\affiliation{%
  \institution{College of Computer Science and Technology, National University of Defense Technology}
  \city{Changsha}
  \state{Hunan}
  \country{China}
}
\email{email_suyi@163.com}

\author{Qian Zhu}
\affiliation{%
  \institution{College of Computer Science and Technology, National University of Defense Technology}
  \city{Changsha}
  \state{Hunan}
  \country{China}
}
\email{zhuqian@126.com}

\author{Yong Dou}
\affiliation{%
  \institution{College of Computer Science and Technology, National University of Defense Technology}
  \city{Changsha}
  \state{Hunan}
  \country{China}
}
\email{yongdou@nudt.edu.cn}

\author{Xinwang Liu}
\affiliation{%
  \institution{College of Computer Science and Technology, National University of Defense Technology}
  \city{Changsha}
  \state{Hunan}
  \country{China}
}

\author{Kele Xu}
\authornote{Corresponding author}
\affiliation{%
  \institution{College of Computer Science and Technology, National University of Defense Technology}
  \city{Changsha}
  \state{Hunan}
  \country{China}
}
\email{kele.xu@ieee.org}

\renewcommand{\shortauthors}{Yulin Sun et al.}

\begin{abstract}
AudioSet is a widely used benchmark in the audio research community and has significantly advanced various audio-related tasks. However, persistent issues with label accuracy and completeness remain critical bottlenecks that limit performance in downstream applications.
To address the aforementioned challenges, we propose a three-stage reannotation framework that harnesses general-purpose audio-language foundation models to systematically improve the label quality of AudioSet. The framework employs a cross-modal prompting strategy, inspired by the concept of prompt chaining, wherein prompts are sequentially composed to execute subtasks (audio comprehension, label synthesis, and semantic alignment). 
Leveraging this framework, we construct a high-quality, structured relabeled version of \textbf{AudioSet-R}. Extensive experiments conducted on representative audio classification models—including AST, PANNs, SSAST, and AudioMAE—consistently demonstrate substantial performance improvements, thereby validating the generalizability and effectiveness of the proposed approach in enhancing label reliability.
The code is publicly available at: \url{https://github.com/colaudiolab/AudioSet-R}.
\end{abstract}



\settopmatter{printfolios=False}

\maketitle

\section{Introduction}

Audio signal processing is a highly interdisciplinary domain encompassing a range of core tasks such as speech recognition~\cite{chen2016efficient,gong2024advanced}, sound event detection~\cite{zhu2020audio,xu2019general}, medical sound analysis~\cite{ji2024weight,bi2025systematic} and emotion recognition~\cite{tu2024higher}—all of which aim to extract discriminative representations from raw audio signals~\cite{deep,survey,quelennec2025masked}. In recent years, the rapid advancement of deep learning has established it as the dominant paradigm in this domain~\cite{sun2023automatic,sun2024self, zhu2025ssast}. In particular, both supervised and self-supervised learning approaches~\cite{xu2023self,xuq2023self,chen2025contrastive} have substantially improved the automation, robustness, and scalability of audio understanding systems, facilitating their deployment across a wide range of real-world applications~\cite{you2022underwater,xu2017north,kershenbaum2025automatic}.

\begin{figure}[!t]
    \centering
    \includegraphics[width=.85\linewidth,keepaspectratio]{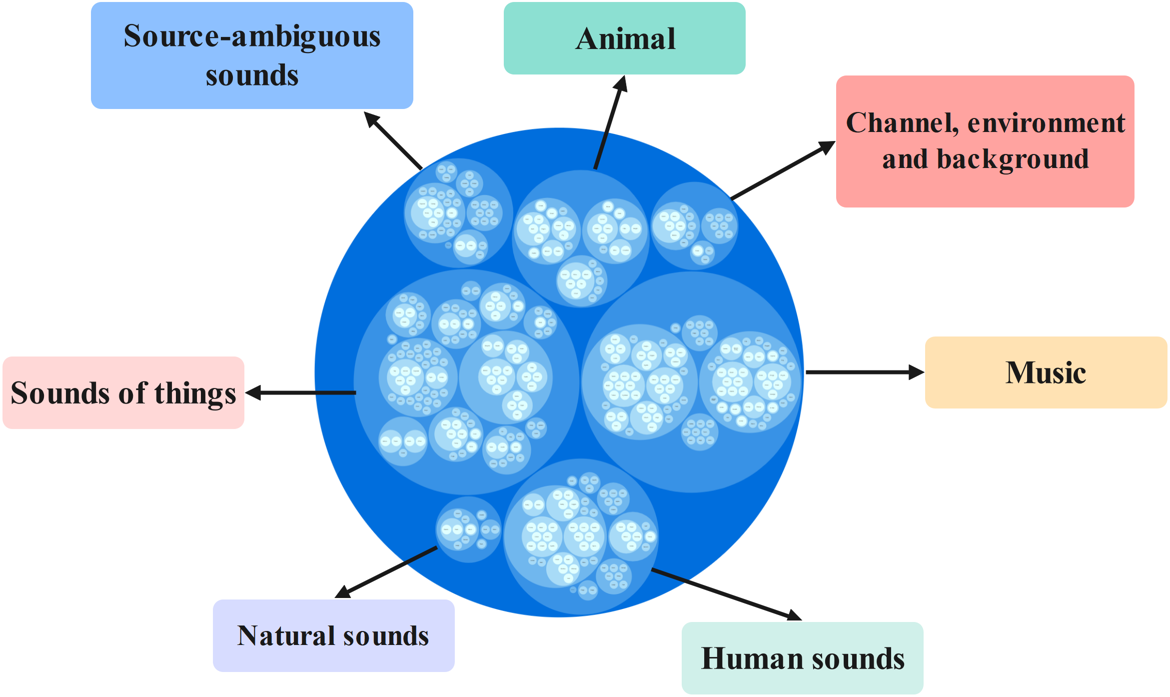}
    \caption{Label Hierarchy Distribution within AudioSet's Seven Top-Level Categories (Limited to Four Levels).}
    \Description{Label hierarchical distribution}
    \label{fig3}
\end{figure}

AudioSet~\cite{audioset}, a large-scale audio dataset released by Google Research, has significantly advanced the development of audio understanding systems. It comprises about 2 million 10-second clips (over 2,000 hours) spanning 527 sound event classes, including human sounds, source-ambiguous sounds, animal, sounds of things, music, natural sounds, and background~\cite{audioset, cnn,zhang2025audio}. The extensive category coverage and large-scale sample size have driven the development of many audio learning frameworks~\cite{schmid2025effective,mulimani2025domain,labbe2024conette}.
For instance, Kong et al.~\cite{kong2020panns} proposed Pretrained Audio Neural Networks (PANNs), CNN-based models trained on AudioSet that significantly improved tagging and various audio recognition tasks. Gong et al.~\cite{gong2021ast} introduced a transformer-based framework using self-attention to capture broader contexts and learn more discriminative features. To tackle AudioSet’s noisy labels, Gong et al.~\cite{gong2022ssast} further developed a self-supervised masked modeling approach that reduces the impact of incorrect annotations.
Despite these advances, two major challenges persist. Supervised methods depend heavily on label quality, which is limited by AudioSet’s weak annotations. Conversely, self-supervised methods do not leverage available hierarchical semantic information, risking the loss of high-level contextual cues. Our analysis of the label depth distribution in AudioSet’s balanced training subset (Figure~\ref{fig3}) shows that most labels lie within the first four hierarchy levels, indicating a dominance of coarse-grained semantics and underutilization of finer-grained distinctions.

In response, several hierarchical refinement strategies have been proposed. Tomasz et al.~\cite{automatic} introduced a two-stage method—Parent-Expanded Labels and Parent-Expanded and Children-Masked Labels—to address semantic redundancy and hierarchical conflicts, thereby improving label balance. Building upon this, Ludovic et al.~\cite{hlp} proposed Hierarchical Label Propagation, a systematic approach that infers missing parent-child annotations using the AudioSet ontology. While these methods achieve certain improvements, they remain inherently dependent on the correctness of the original labels, and their performance is significantly constrained by label noise and incompleteness.

We propose an automated relabeling framework that leverages state-of-the-art audio-language foundation models in conjunction with a Prompt Chaining strategy. Specifically, the Qwen-Audio multimodal model is first employed to extract rich semantic information from each audio clip by generating a structured three-part textual description encompassing the primary content, presence of human voice, and presence of music. This textual summary is then processed by the Mistral language model, which utilizes structured prompts to generate semantically precise and contextually grounded label predictions. Finally, the DeepSeek R1 model aligns the predicted labels with the predefined 527 AudioSet categories through exact matching and synonym recognition, thereby achieving consistent label mapping and semantic normalization.
The main contributions of this work are summarized as follows:
\begin{itemize}
    \item We propose a novel three-stage relabeling framework that integrates audio-language foundation models with a Prompt Chaining strategy, enabling structured semantic extraction, label prediction, and ontology-based alignment for weakly labeled audio data.
    \item We construct a high-quality relabeled version of AudioSet (\textbf{AudioSet-R}), leveraging multimodal and large language models to enhance label accuracy, completeness, and semantic consistency—addressing key limitations in both supervised and self-supervised paradigms.
    \item Extensive experiments on benchmark audio classification models (e.g., AST, PANNs, SSAST, AudioMAE), demonstrate the generality and effectiveness of our relabeling approach, with consistent performance improvements observed across both supervised and semi-supervised learning settings.
\end{itemize}

\section{Related Work}{\label{relatedwork}}
AudioSet is a widely used large-scale dataset in the audio domain that has substantially contributed to advancements in various audio-related tasks~\cite{sun2024automated,mulimani2024class,zhou2025dense,choudhury2025rejepa}. Nevertheless, as a weakly labeled dataset, its inherent label quality limitations significantly constrain the performance of downstream applications~\cite{cai2024mat,suryawanshi2025audio,shah2024importance}. To address these challenges, prior research has explored hierarchical label propagation techniques aimed at mitigating issues such as label redundancy, missing annotations, and semantic inconsistency. 
For instance, Grzywalski et al.~\cite{automatic} proposed two label correction strategies: Parent-Expanded Labels (PEL) and Parent-Expanded and Children-Masked Labels (PE-CML). It propagate child labels upward to their respective parent labels during model training to enforce hierarchical consistency. Simultaneously, when a parent label is present, corresponding child labels are masked in the loss function to alleviate the negative impact of label noise. These strategies effectively reduce semantic redundancy and hierarchical conflicts, improving feature stability and model learning in multi-label classification.
Extending this line of work, Tuncay et al.~\cite{hlp} introduced Hierarchical Label Propagation (HLP), a more systematic label enhancement approach designed to address missing parent-child labels within the AudioSet ontology. HLP propagates annotated positive labels upward along semantic paths to all ancestor classes, thereby constructing a more complete and hierarchically consistent label hierarchy. 

Despite these advances, the effectiveness of hierarchical propagation techniques fundamentally depends on the accuracy of the original labels, limiting their capacity to fully address more complex label quality issues such as mislabeling, missing annotations, and redundancy~\cite{Dinkel2022pseudo}. Erroneous original labels may be propagated and even amplified through hierarchical mechanisms, thereby degrading overall label accuracy. Moreover, these methods exhibit limited capability in handling semantic ambiguity, incomplete label coverage, and polysemy, which constrains their robustness and generalizability in large-scale audio labeling systems~\cite{gao2022multi}. Consequently, there is an urgent need to develop large-model-assisted relabeling frameworks that incorporate semantic understanding and cognitive reasoning, aiming to fundamentally improve the accuracy, consistency, and scalability of the AudioSet labeling system.

\begin{figure*}[!t]
    \centering
    \includegraphics[width=1\linewidth]{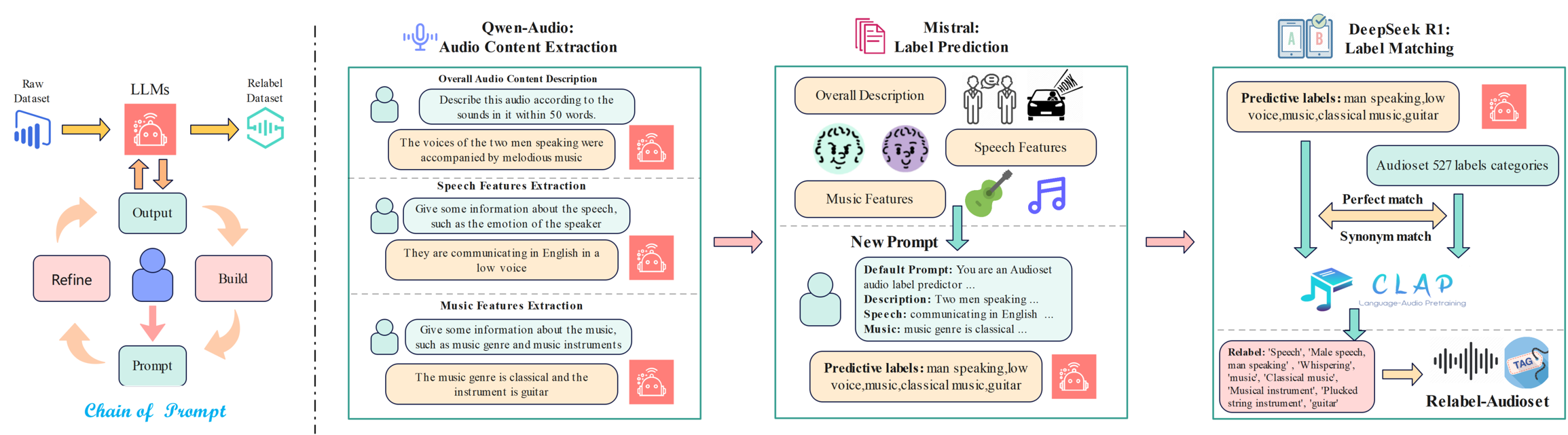}
    \caption{Illustrates the proposed three-stage relabeling framework for AudioSet. The left part presents the overall concept of multi-round prompt chaining, where each model's output is incorporated into a fixed instruction template to construct the next prompt, progressively guiding the label generation process. The right part details the implementation pipeline, sequentially integrating Qwen-Audio for fine-grained semantic parsing, Mistral for initial label prediction, DeepSeek R1 for taxonomy alignment, and CLAP for semantic similarity filtering, ultimately producing a high-quality relabeled tag set.}
    \Description{The relabeling framework.}
    \label{fig1}
\end{figure*}

\section{Methodology}

We propose a Prompt Chaining-based relabeling framework in which three large language models (LLMs) collaboratively refine labels for AudioSet audio clips~\cite{prompt, bai2024audiosetcaps,li2023multi}. Prompt Chaining~\cite{chainofthought,wu2022aichains, wu2022promptchainer} is a structured prompting strategy that decomposes complex tasks into sequential subtasks, enabling intermediate reasoning, stable outputs, and context-aware decision-making~\cite{chain, engineer}. Motivated by these strengths, we design a three-stage relabeling pipeline aligned with the Prompt Chaining paradigm.

As illustrated in Figure~\ref{fig1}, the process begins with Qwen-Audio~\cite{chu2023qwen,wang2025enabling,zhou2025advancing}, which performs semantic decomposition by generating structured textual summaries from raw audio, covering core aspects such as primary content, human voice presence, and musical elements. Next, Mistral leverages these structured summaries within prompt templates to produce semantically precise and contextually relevant candidate labels. Finally, DeepSeek R1~\cite{guo2025deepseek,hayder2025highlighting,chen2025suitability} aligns the predicted labels with the 527 AudioSet categories using exact and synonym-aware matching, ensuring consistency and taxonomic compliance.

This multi-model chained framework offers a scalable and automated solution to audio annotation. It enhances label completeness and semantic accuracy while effectively addressing the subjectivity, noise, and inconsistency issues often found in manual labeling or single-pass prompting approaches.

\subsection{Audio Content Extraction}

Accurate audio tag prediction relies on obtaining fine-grained, high-quality semantic descriptions that capture the core content of each audio segment~\cite{you2022masked,chinthalapani2025audiocast,lai2024multichannel}. These descriptions serve as the foundational input for downstream label prediction, enhancing both semantic interpretability and consistency. To this end, we adopt Qwen-Audio as the core component of the semantic extraction module, based on its demonstrated effectiveness in audio understanding and cross-modal generation~\cite{10852359, 10887618,ghosh2024gama}.

In our framework, we utilize the `model.chat' interface and design a multi-turn prompting strategy to extract structured semantic representations from raw audio inputs. Prompt templates are crafted to regulate both the semantic scope and the format of each response. The design of this three-round prompting mechanism is motivated by the hierarchical category topology of AudioSet. For example, within the parent category of Human sounds, subcategories include man speaking, baby crying, and woman laughing; similarly, Musical instruments includes classes such as piano and guitar. Capturing such fine-grained distinctions requires explicit semantic decomposition at multiple levels. Specifically:
Round 1 prompts the model to generate a concise ($\leq$50 words) holistic description of the audio.
Round 2 (if human voice is present) requires identification of speaker-related features such as emotional state, gender, and spoken language.
Round 3 (if music is present) instructs the model to specify the genre and instruments involved.
The outputs from all turns are aggregated into a standardized, three-part textual summary encompassing: (1) primary content, (2) vocal information, and (3) musical content. This structured representation supports hierarchical semantic organization and modular downstream processing~\cite{10852359, 10731549}.
To improve robustness, we implement a fault-tolerant retry mechanism allowing each audio sample up to Qwen-try-num generation attempts. This ensures compliance with output length and formatting constraints, or otherwise terminates upon reaching the maximum retry limit~\cite{bai2024audiosetcaps}.
This strategy enables Qwen-Audio to reliably extract semantically rich, structurally consistent descriptions from complex auditory scenes, thereby provides high-quality input for label prediction module.

\subsection{Label Prediction}

In examining the limitations of direct audio tagging, we observed that relying solely on high-level semantic descriptions without explicit control often led to over-labeling or inconsistent tag quality—particularly in cases where fine-grained semantic distinctions were subtle but crucial (such as \textit{crying and sobbing sounds}). Moreover, analysis of the AudioSet corpus revealed that the average number of valid labels per clip is relatively low (2.39 in training, 2.55 in evaluation), indicating a need for restraint and contextual precision in tag generation.

To this end, we designed a context-aware tagging strategy based on Mistral, a large language model developed by Mistral AI~\cite{10434081}. This strategy leverages the structured three-part descriptions from Qwen-Audio as input and embeds them into semantically enriched prompt templates tailored for label generation. Informed by corpus statistics, we introduced a label count control mechanism within the prompt design, enabling the model to maintain a balance between semantic coverage and output brevity.
Through this strategy, Mistral performs guided, inference-based prediction that is logically consistent and semantically grounded. By combining detailed contextual input with instructional constraints~\cite{10776884}, the model achieves improved control over label validity while ensuring computational efficiency—demonstrating its practicality in scalable and automated audio relabeling tasks.

\subsection{Label Matching}

Preliminary analysis revealed that many tags predicted in Stage 2, though semantically valid, did not conform to the standardized AudioSet ontology. These free-form labels varied in expression, granularity, and wording, leading to semantic misalignment and hindering downstream integration.
To resolve this, we designed a label alignment strategy using DeepSeek R1~\cite{lu2024deepseek}, a multimodal LLM. The goal is to map non-standard tags to the 527 AudioSet categories through a three-tier matching mechanism: (1) Exact Matching for direct correspondence, (2) Fuzzy Matching to tolerate lexical variations, and (3) Synonym Matching using semantic resources~\cite{barbany2024leveraging}.

While the alignment strategy ensured structural consistency, we observed that some mapped labels lacked semantic relevance to the corresponding audio. This revealed a need for post-alignment filtering based on content-label consistency.
To address this, we integrated CLAP~\cite{wu2023large} to evaluate semantic alignment. For each candidate label, we compute its CLAP similarity score with the audio and retain only those surpassing a predefined threshold~\cite{10095889}. This filtering step improves label reliability by removing low-confidence or semantically misaligned predictions. This process can be formulated as follows:
\begin{align}
    \text{S}(x, y) = \frac{f_a(x) \cdot f_t(y)}{\|f_a(x)\| \cdot \|f_t(y)\|}
\end{align}
where $x$, $y$ denote the audio sample and corresponding text, $S(x, y)$ represents their similarity score, and $f_a$, $f_b$ are the audio and text encoders, respectively. These two strategies, ontology alignment with DeepSeek R1 and semantic filtering with CLAP, effectively address label inconsistency and weak semantic grounding. Integrated with earlier stages (Qwen-Audio for description, Mistral for prediction), they form a robust, modular relabeling pipeline.

\section{Experiments}
\subsection{Experimental Setup}
\textbf{Training configurations:}To evaluate the effectiveness of the proposed \textbf{AudioSet-R}, we conducted a series of experiments using a range of mainstream audio classification models under both supervised and self-supervised learning paradigms. 
The training configuration includes the following model architectures:

\begin{itemize}
    \item \textbf{Fine-tuning of self-supervised models:} We selected two representative self-supervised pretrained models, SSAST~\cite{gong2022ssast} and AudioMAE~\cite{NEURIPS2022_b89d5e20}, for downstream fine-tuning on \textbf{AudioSet-R}.
    \item \textbf{Training of supervised models:} We employed AST~\cite{gong2021ast} as well as PANNs (CNN6 and CNN14)~\cite{kong2020panns} to evaluate performance under fully supervised settings.
\end{itemize}

\textbf{Network Architectures:} we select two representative methodologies encompassing both supervised and self-supervised learning paradigms. In the supervised learning category, we employ the PANNs family (CNN-based architectures) and the Audio Spectrogram Transformer (AST), which utilizes a transformer-based architecture. Both models are trained from scratch and heavily rely on high-quality, human-annotated labels to achieve optimal performance.
In the self-supervised learning paradigm, we adopt two transformer-based models: SSAST and AudioMAE . These models leverage large-scale unlabeled audio datasets for pretraining, enabling them to learn generalizable audio representations. Subsequently, they are fine-tuned on downstream tasks with limited annotated data, demonstrating effective adaptation and improved performance.

\textbf{Evaluation Metrics:} For evaluation, we adopt mean Average Precision (mAP), a widely recognized and representative metric for multi-label classification tasks~\cite{chen2022beats}. mAP is calculated by first computing the Average Precision (AP) for each class and then averaging these values across all classes, thereby providing a comprehensive assessment of model performance, especially in scenarios with imbalanced label distributions.

Unlike traditional metrics such as accuracy or recall, which can be biased in the presence of label imbalance, mAP offers a more robust and informative evaluation. It effectively reflects both the discriminative ability and prediction stability of the model, making it particularly suitable for multi-label classification in our study. Consequently, mAP is employed as the primary evaluation metric throughout all experiments.

Specifically, for each category $c$, its average Precision $AP_c$ is defined as the area integral under its precision-Recall curve, formalized as:
\begin{equation}
    \text{AP}_c = \int_{0}^{1} p_c(r) \, dr
\end{equation}
where $p_c(r)$ represents the accuracy of category $c$ under the recall rate $r$. The final mAP is the average of the aps of all $C$ categories, that is:
\begin{equation}
    \text{mAP} = \frac{1}{C} \sum_{c=1}^{C} \text{AP}_c
\end{equation}

\textbf{Training Details:} In this study, we adopt the experimental setup described in the original AST paper, utilizing a DeiT-based variant. The model processes spectrogram inputs through multiple stacked Transformer encoder layers, followed by a LayerNorm operation and a linear classification head for multi-label prediction. Training is conducted from scratch over 25 epochs, employing a learning rate of 0.00001 and a batch size of 16.

\subsection{Quantitative Comparison}

To comprehensively evaluate the effectiveness of the proposed \textbf{AudioSet-R}, we conducted extensive experiments on multiple mainstream audio classification models encompassing both supervised and self-supervised learning paradigms~\cite{gong2022ssast, superb}.

The AST model, trained and evaluated on the relabeled \textbf{AudioSet-R}, achieved a mAP of 0.1310, representing a significant improvement of approximately 3.2 percentage points over the baseline model trained and evaluated on the original AudioSet (mAP = 0.0979). When trained on the original training set but evaluated on the relabeled validation set, the AST model’s performance improved by roughly 1 percentage point, indicating the superior label quality of the relabeled validation set and its effectiveness in addressing label omissions and inaccuracies. In contrast, training on the relabeled set while evaluating on the original validation set resulted in a decrease in mAP. This decline is attributed to annotation inconsistencies in the original validation set, which impeded reliable feedback despite correct predictions by the model.

Within the PANNs framework~\cite{kong2020panns}, CNN6 and CNN14 were selected as representative convolutional architectures featuring 6 and 14 layers, respectively. The deeper CNN14 exhibits enhanced feature representation capacity. Both models were independently trained on the original and relabeled balanced training subsets and evaluated on the corresponding validation sets. Results indicate consistent performance improvements when using the relabeled annotations, with mAP values increasing to 0.2620 for CNN6 and 0.2690 for CNN14, substantially exceeding the baselines trained with original labels.
Collectively, these results demonstrate that the refined \textbf{AudioSet-R} annotations contribute to more robust and generalizable model performance across diverse architectures. The observed improvements in both Transformer-based and CNN-based models confirm the efficacy and wide applicability of the proposed relabeling approach in multi-label audio classification tasks.

\begin{table}[!t]
\centering
\caption{Comparison of training and fine-tuning performance on \textbf{AudioSet-R} across various supervised and self-supervised audio classification models.}
\resizebox{1.0\linewidth}{!} {
\begin{tabular}{cccccc}
\hline
\multirow{2}{*}{Model} & \multirow{2}{*}{Venue} & \multirow{2}{*}{\#Params(M)} & \multirow{2}{*}{Train Data} & \multicolumn{2}{c}{Eval Data} \\ \cline{5-6} 
                       &         &                              &                             & Original Eval   & Relabel Eval   \\ \hline
\multirow{2}{*}{SSAST}   & \multirow{2}{*}{AAAI 2022}  & \multirow{2}{*}{89.0}   & Original Train              & 0.2528          & 0.2603        \\
                        &        &                              & Relabel Train                & 0.2472          & \bfseries 0.2989        \\
\multirow{2}{*}{AudioMAE} & \multirow{2}{*}{NeurIPS 2022}  & \multirow{2}{*}{86.0}   & Original Train              & 0.3408          & 0.3123        \\
                       &        &                              & Relabel Train                & 0.2917          & \bfseries 0.3425        \\ \hline
\multirow{2}{*}{AST} & \multirow{2}{*}{Interspeech 2021}       & \multirow{2}{*}{87.0}          & Original Train              & 0.0979          & 0.1078        \\
                        &        &                              & Relabel Train                & 0.0936          & \bfseries 0.1310         \\
\multirow{2}{*}{CNN6}  &   \multirow{2}{*}{TASLP 2020}      & \multirow{2}{*}{4.8}         & Original Train              & 0.2530           & 0.2310         \\
                       &         &                              & Relabel Train                & 0.2140           & \bfseries 0.2620         \\
\multirow{2}{*}{CNN14} &    \multirow{2}{*}{TASLP 2020}     & \multirow{2}{*}{79.6}        & Original Train              & 0.2630           &  0.2340             \\
                        &        &                              & Relabel Train                & 0.2230           & \bfseries 0.2690             \\ \hline
\end{tabular}
}
\label{table2}
\end{table}

For self-supervised learning models, results demonstrate that fine-tuning on the relabeled training set and evaluating on the corresponding relabeled validation set yields the best performance, with an mAP of 0.2989—substantially surpassing the baseline performance using the original label set (mAP = 0.2528). Notably, even when fine-tuned on the original training set but evaluated on the relabeled validation set, the model achieves a ~1 percentage point improvement in mAP. This confirms the enhanced label quality, consistency, and semantic granularity of the relabeled validation set. Conversely, when fine-tuned on the relabeled training set but evaluated on the original validation set, performance deteriorates, reflecting the adverse impact of incomplete or inaccurate annotations in the original set that obscure correct predictions and compromise evaluation reliability.
\begin{figure}[!t]
    \centering
    \includegraphics[width=1\linewidth,keepaspectratio]{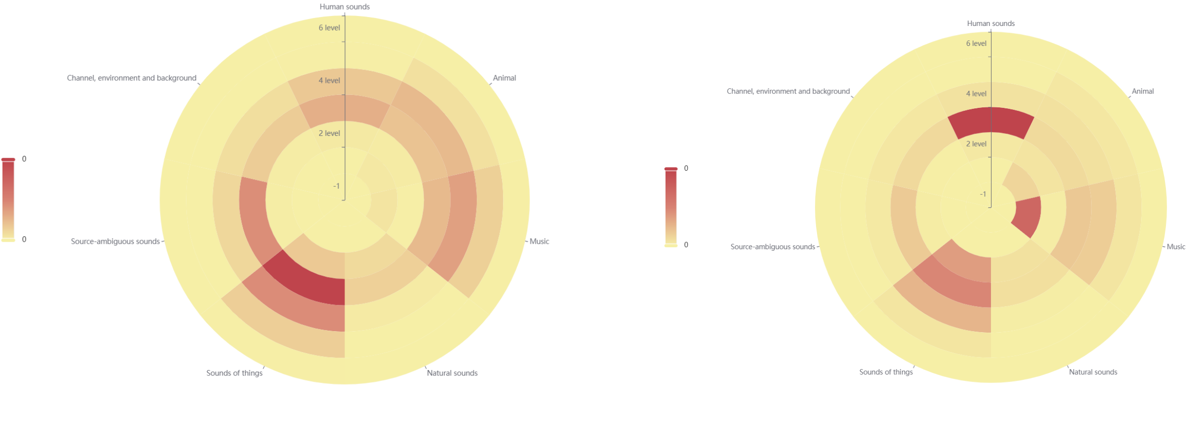}
    \caption{Distribution of ground-truth and predicted labels for misclassified samples.}
    \Description{Distribution comparison chart}
    \label{fig4}
\end{figure}

\begin{figure}[t]
    \centering
    \includegraphics[width=0.8\linewidth,keepaspectratio]{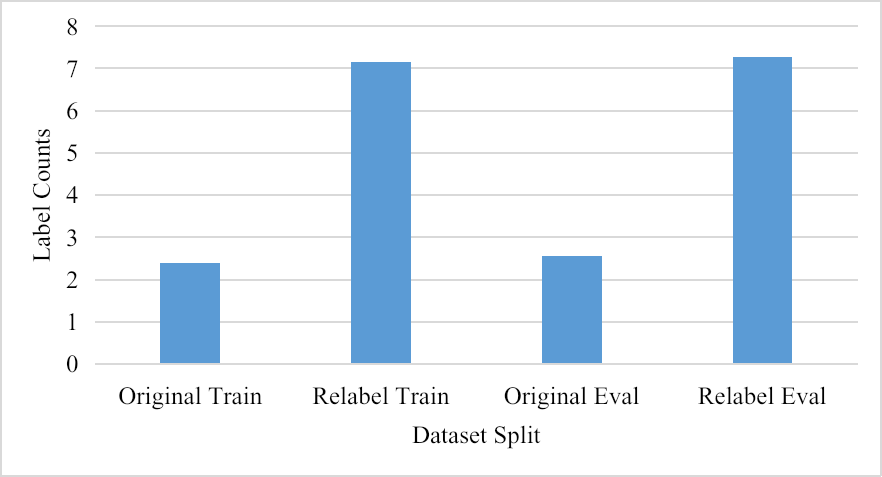}
    \caption{The average number of labels in the dataset.}
    \Description{Distribution comparison chart2}
    \label{fig:label_count}
\end{figure}

\begin{figure*}[t]
    \centering
    \includegraphics[width=1\linewidth,keepaspectratio]{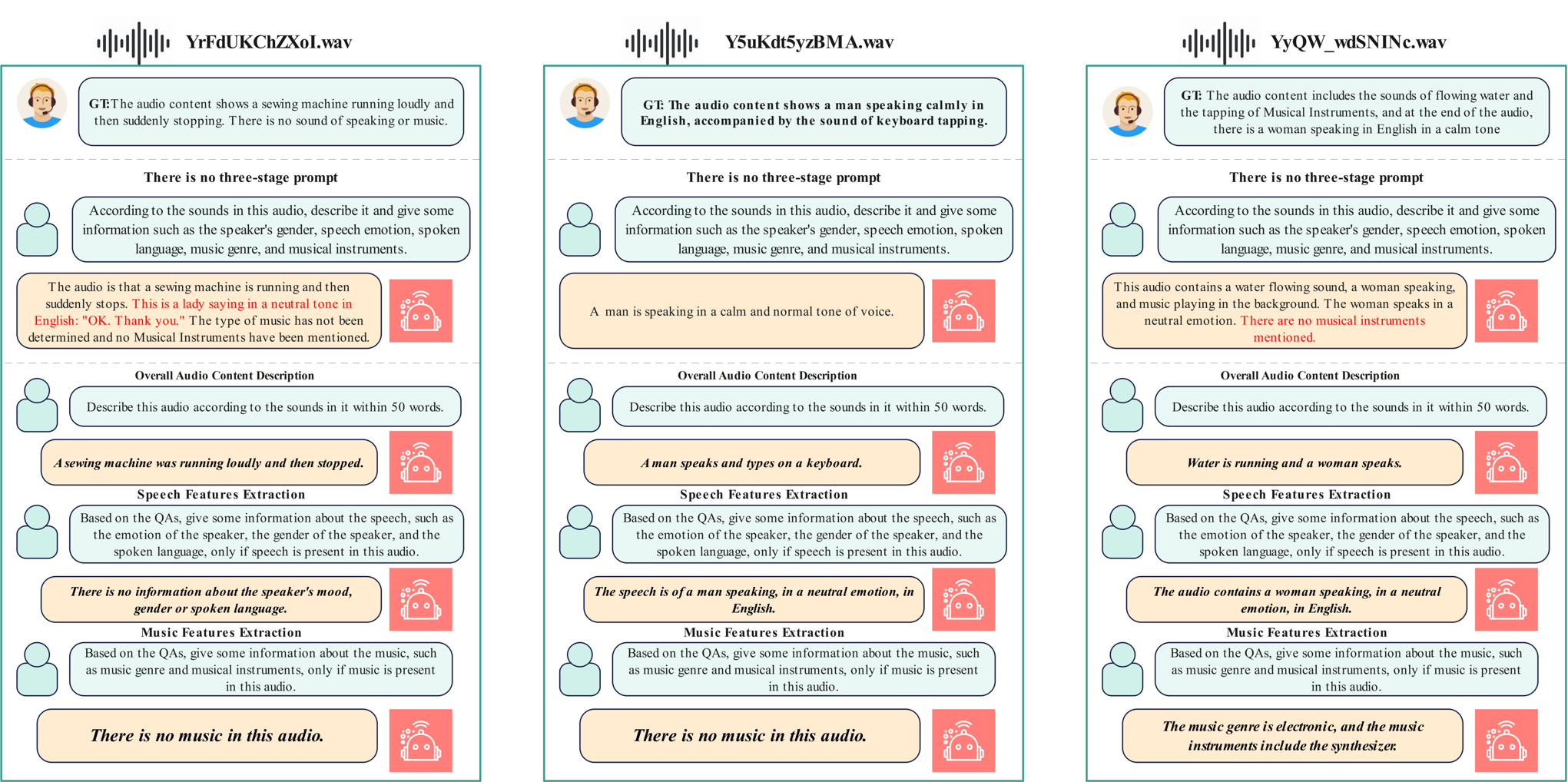}
    \caption{The detailed analysis for three-round audio content extraction. }
    \Description{Distribution comparison chart3}
    \label{fig:case}
\end{figure*}

AudioMAE, a self-supervised architecture adapted from the MAE originally designed for vision tasks, learns context-aware audio representations by reconstructing masked time-frequency patches~\cite{NEURIPS2022_b89d5e20}. In our experiments, we fine-tune a pretrained AudioMAE model using a learning rate of 1e-3 for 60 epochs, while keeping other hyperparameters unchanged. Results indicate consistent improvements when trained and evaluated on the relabeled dataset, reaffirming that AudioMAE benefits from higher-quality annotations and exhibits sensitivity to improved label fidelity.

In addition, as shown in Figure~\ref{fig4}, an error analysis of misclassified validation samples reveals three primary sources of prediction error: (1) confusion among fine-grained labels—particularly within categories such as ``Human sounds''; (2) misclassification of environmental sounds into semantically similar classes, e.g., confusing ambient noise with `Animal' labels; and (3) incorrect predictions on acoustically ambiguous or indistinct audio segments. These issues underscore the challenges posed by subtle inter-class semantic differences and ambiguous label boundaries.

It is also observed that AudioMAE, while effective in capturing structural audio features, is not explicitly optimized for discriminating fine-grained semantic categories. This inherent limitation reduces its sensitivity to nuanced label distinctions. We hypothesize that integrating more discriminative downstream classifiers or task-specific adaptation strategies may further enhance its classification accuracy and label resolution capacity.

In summary, the experimental results provide compelling evidence that the \textbf{AudioSet-R} substantially enhances label quality in both the balanced training and validation subsets. These improvements consistently translate into measurable performance gains across a variety of supervised and self-supervised models during both training and fine-tuning stages. The empirical outcomes affirm not only the effectiveness of the proposed relabeling strategy but also its robustness and generalizability in multi-label audio classification tasks. Moreover, these findings establish a strong foundation for extending the relabeling methodology to the unbalanced training subset.
Importantly, the results underscore the pivotal role of high-quality labels in advancing model performance for complex audio understanding. This highlights the practical relevance and theoretical significance of label refinement as a key enabler of more accurate, robust, and generalizable audio representation learning.

\subsection{Statistical Analysis of Label Counts}
To verify the semantic richness and information density of the \textbf{AudioSet-R}, we performed a statistical analysis of label counts, as the number of labels per audio sample serves as a key indicator of semantic coverage and labeling granularity.
As shown in Figure~\ref{fig:label_count}, the proposed relabeling strategy yields a marked increase in the average number of labels per clip across both the balanced training and evaluation subsets, compared to the original AudioSet annotations. This suggests improved annotation density, with more relevant audio concepts accurately captured.

\subsection{Further Analysis}
To provide a detailed illustration of the proposed three-round audio content extraction strategy, we conduct a focused case study. Specifically, our analysis of the AudioSet ontology reveals that labeling errors are most prevalent in major categories with deep and complex parent–child hierarchies—particularly Human sounds and Musical instruments. These hierarchical structures often introduce semantic ambiguity, making fine-grained label assignment challenging. To examine the effectiveness of our method in addressing these issues, we randomly selected audio samples from these two categories. Ground-truth annotations were established through expert listening and manual transcription. We then compared the label outputs generated by a single-pass extraction approach with those obtained using our three-round strategy. As shown in Figure~\ref{fig:case}, the single-pass method frequently results in erroneous or overly generic labels, often failing to reflect subtle semantic distinctions. In contrast, the proposed three-round strategy significantly reduces such errors, yielding more accurate and semantically consistent label assignments.

\section{Conclusion}

This study presents a three-stage label relabeling framework that leverages the complementary strengths of multiple large-scale audio-language and vision-language models to enhance the accuracy, completeness, and semantic granularity of AudioSet—a widely adopted benchmark for audio understanding. The proposed pipeline integrates Qwen-Audio for fine-grained audio captioning, Mistral for prompt-based label generation, and DeepSeek R1 for ontology alignment, while CLAP is used to evaluate semantic alignment between audio content and candidate labels for final refinement.
The resulting \textbf{AudioSet-R} exhibits significantly improved label density and consistency. Comprehensive experiments across diverse model architectures—including AST, PANNs, SSAST, and AudioMAE, demonstrate consistent and substantial performance gains in multi-label audio classification. These findings not only validate the effectiveness and generalizability of the relabeling strategy but also highlight the promise of large-scale multimodal models in enhancing weakly labeled datasets for complex audio analysis tasks.

\begin{acks}
This work is supported by National Science and Technology Major Project (2023ZD0121101), National University of Defense Technology (ZZCX-ZZGC-01-04) and Major Fundamental Research Project of Hunan Province (2025JC0005).
\end{acks}

\bibliographystyle{ACM-Reference-Format}
\bibliography{ref}

\end{sloppypar}
\end{document}